\documentstyle[12pt,twoside,epsfig,espcrc1]{article}
 \input epsf.def
 \epsfverbosetrue

\let\hat\widehat 
\let\bar\overline 
\def\hhhh{ \hat{\bar{\psi}}\,\hat{\bar{\psi}}\, \hat{\psi}\, \hat{\psi}}
\def\pppp{ \bar{\psi} \,\bar{\psi}\, \psi\, \psi}
\def\bp{ \bar{\psi}\, \psi}
\def\sq{\mbox{\scriptsize$\sqcap$\llap{$\sqcup$}}}

\def\DD{{\cal D}}
\def\RT{\mathop{\rm Re \, Tr}}

\def\notp{/ \llap{$p$}}
\def\notA{/ \llap{$A$}}
\def\notpt{/ \llap{$\partial$}}

\title{Insight into the Role of Instantons and their Zero Modes\\ from Lattice QCD}

\author{J.W. Negele\address{Center for Theoretical Physics\\ 
Laboratory for Nuclear Science
and Department of Physics\\ 
Massachusetts Institute of Technology, Cambridge
MA 02139\\
}\thanks{This work is
supported in part by funds provided by the US Department of Energy (DOE) under
cooperative research agreement \#DF-FC02-94ER40818.}}

\begin{document}
\maketitle

\begin{abstract}

Evidence from lattice QCD calculations is presented showing that instantons and
their associated zero modes play a major role in the physics of light hadrons and
the propagation of light quarks in the QCD vacuum.

\end{abstract}

\section{Introduction and Motivation}

During the quarter of a century I have known Koichi Yazaki as a colleague,
collaborator and friend, we have shared an abiding interest in many body physics
- understanding how strongly interacting systems develop
their rich and complex properties from their fundamental underlying
interactions.  Hence, it is a special pleasure and privilege to be able to take part
in this symposium being held in his honor and to talk about insight my
collaborators and I have gotten from lattice QCD into the role of instantons and
their associated zero modes in the physics of hadrons.

One of the great advantages of the path integral formulation of quantum
mechanics and field theory is the possibility of identifying
non-perturbatively the stationary configurations that dominate the action
and thereby identify and understand the essential physics of complex
systems with many degrees of freedom.  Thus, the discovery of instantons
in 1975~\cite{belavin+75} gave rise to great
excitement and optimism that they were the key to understanding QCD. 
Indeed, in contrast to other many body systems in which the quanta
exchanged between interacting fermions can be subsumed into a potential, it
appeared that QCD was fundamentally different, with topological
excitations of the gluon field dominating the physics and being
responsible for a host of novel and important effects including the
$\theta$ vacuum, the axial anomaly, fermion zero modes, the mass of the
$\eta^\prime$, and the chiral condensate.  However, despite nearly a
quarter of a century of theoretical effort, it has
not been possible to proceed analytically beyond the dilute instanton gas
approximation~\cite{callan+78a}. In the intervening years, the  
instanton liquid  model~\cite{shuryak82,diakonov+86,schafer+98} provided a
successful phenomenology and qualitative physical understanding, but a 
quantitative exploration of the role of instantons in nonperturbative QCD
has had to wait until lattice QCD became sufficiently sophisticated and
sufficient resources could be devoted to the study of instanton physics. 

Our basic strategy will be to reverse the usual analytical process of calculating
stationary configurations and approximately summing the fluctuations around
them. Rather, we will use Monte Carlo sampling of the path integral for QCD
on a lattice to identify typical paths contributing to the action and then work
backwards to identify the smooth classical solutions about which these paths are
fluctuating. The physical picture that arises  corresponds closely to
the physical arguments and instanton models of Shuryak and
Diakonov\cite{shuryak82,diakonov+86,schafer+98} in which the zero modes
associated with instantons produce localized quark states, and quark propagation
proceeds primarily by hopping between these states.  

\subsection{Aspects of continuum instanton physics}

To put lattice investigations in context, it is useful to recall relevant
aspects of continuum instanton physics. Working in Euclidean time, we
evaluate a path integral of the form $\int{\cal D}[A]e^{\int d^4x\, S[A]}$.
Hence, as in statistical mechanics, the weight of a configuration depends
not only on its energy, but also on its entropy -- 
 the number of ways it
can be realized. In addition, tunneling solutions arise as periodic
classical solutions in an inverted potential. Hence, we may expect BPST
instantons~\cite{belavin+75} which connect degenerate minima of differing
winding number with the self-dual gauge potential
$A_{\mu}^a(x) = {{2 \eta_{a\mu\nu}x_{\nu}}\over{x^2 + \rho^2}}$ having
scale-invariant action $S = {1\over 4}\int d^4x\, F_{\mu\nu}^a
F_{\mu\nu}^a = {8\pi^2 g^{-2}} \equiv S_0$ to have a significant
presence in the vacuum due to the high entropy associated with translation,
size, and color orientation. The action and topological charge density are
localized around the center as $\pm F_{\mu\nu}^a \tilde F_{\mu\nu}^a =
F_{\mu\nu}^aF_{\mu\nu}^a = {192\rho^4\over (x^2+\rho^2)^4}$\ . The
$dn_I \sim ({8\pi^2\over g^2})^{\scriptscriptstyle 2N_c}{d\rho\over
\rho^5} d^4x^{(I)}  exp\{{-{8\pi^2\over g^2(\Lambda^{-1})}}\} (\Lambda \rho)^{11
N_c\over 3} $, where the prefactor
and the running of the coupling constant in the last factor  produce a distribution
of instantons $\sim \rho^6$ for SU(3).  Physically, we expect this distribution to be
cut off at large
$\rho$ by interactions between instantons and by fluctuations when the amplitude
of a sufficiently large instanton becomes small relative to  quantum fluctuations.
It is the difficulty in treating these infrared effects that has stymied analytic
progress.

From the axial anomaly, 
$\partial_{\mu}\sum_f\bar\psi\gamma_{\mu}\gamma_{5}\psi =$
$2m\sum_f\bar\psi
\gamma_5\psi + {N_f\over 16 \pi^2}F_{\mu\nu}^a \tilde F_{\mu\nu}^a$,
the topological charge satisfies the index theorem and, for periodic
systems, may be expressed in terms of fermion eigenfunctions 
$Q\!=\!{g^2\over
32\pi^2}\int F\tilde F
=n_L-n_R=m\sum_{\lambda}{{\int \psi_{\lambda}^{\dag}(x)
\gamma_5\psi_{\lambda}(x)\over m+i\lambda}} $,
 where $n_L$ and $n_R$ denote the number of fermion zero
modes.  For an isolated instanton, the zero mode is 
$\psi_0(x)={{\rho\gamma\cdot\hat
x(1+\gamma_5)}\over2\pi(x^2+\rho^2)^{3/2}}\phi$. In the limit of light quarks,
the Greens function for N$_f$ quarks reduces to the product of zero modes 
$\prod_f \det[ {D\!\!\!\!/}+m] \bar\psi_f(x)\psi_f(y)
{\;\longrightarrow _{\!\!\!\!\!\!\!\!\!\!\!\!\!\scriptscriptstyle {
\atop m\rightarrow 0}}}
\prod_f \psi_0^{ }(x) \psi_0^{\dag}(y)$ and gives rise to the 't Hooft
interaction. Thus, light quarks propagate by zero modes which
in turn arise from instantons.  Based on large N arguments, the
Veneziano-Witten formula~\cite{witten79,veneziano79} relates the
$\eta^{\prime}$ mass to the topological susceptibility in the pure gluon
sector\\
$\chi \equiv \int {d^4x\over V} \langle Q(x)Q(0)\rangle  = 
{f_{\pi}^2\over2N_f}(m_{\eta}^2 + m_{\eta^{\prime}}^2 -2m_K^2)$
yielding the expectation that $\chi = (180\, MeV)^4$. 

The instanton liquid model~\cite{shuryak82,diakonov+86,schafer+98} provides
an economical phenomenology of instanton mediated quark propagation in the
QCD vacuum. The integral over all gluon fields that one evaluates in lattice 
QCD using an ensemble of configurations sampling the action is replaced by
 an
ensemble of instantons and antiinstantons 
 of size $\rho \sim 1/3$\,fm and
density $n\sim 1\,\mbox{fm}^4$ randomly distributed in space
and color orientation,  where the values of $\rho$ and $n$ are determined
from the physical gluon and chiral condensates. 
One may think of the 't Hooft interaction as a vertex that absorbs
left-handed particles of each flavor and creates corresponding
right-handed particles, and {\it vice versa} for antiinstantons. Mesons
then propagate in the QCD vacuum by the hopping of quark-antiquark
pairs between these vertices, and the qualitative features of the
channel dependence arises naturally with the pion channel being strongly attractive,
the scalar channel repulsive, and the interaction in the rho channel very weak.
The chiral condensate arises
naturally in this picture by the fact that the zero modes for isolated instantons
mix in the instanton liquid giving rise to a finite density of states at low
virtuality.   

\subsection{Lattice QCD}
A QCD observable is evaluated by defining quark and gluon
variables on the sites and links of a space-time lattice, writing a Euclidean path
integral of the generic form\cite{negele98a} 
$ \langle Te^{-B\hat{H}} \hhhh \rangle 
= Z^{-1} \int  \DD(U) \DD (\bp) e^{-\bar{\psi} M(U) \psi - S(U)} \pppp
\label{E:JN:1}  \nonumber\\
 = Z^{-1} \int  \DD(U) e^{\ln \det M(U) - S(U)} M^{-1}
(U) M^{-1} (U)
$
%
%
and evaluating the final integral over gluon link variables $U$ using the
Monte Carlo method.  The link variable is $U=e^{iagA_\mu(x)}$, the Wilson
gluon action is $S(U)=
\frac{2n}{g^2} \sum_{\sq} (1-\frac{1}{N} \RT U_{\sq})$ where
$U_{\sq}$ denotes the product of link variables around a single
plaquette, and $M (U)$ denotes the discrete Wilson approximation to the
inverse propagator $M (U) \to m+ \notpt  + \ ig \, \notA$.

%
%

Vacuum correlation functions for space-like separated hadron
currents calculated in lattice QCD display the qualitative
behavior expected from the 't Hooft interaction and agree
semi-quantitatively with the instanton liquid model. As emphasized in
ref~\cite{shuryak93}, correlation functions of the form $R(x) \equiv \langle 0|T
J_{\mu}(x) J_{\mu}(0)|0\rangle $ characterize the spatial and channel dependence
of the interaction between quarks and antiquarks and thus supplement hadron
bound state properties like phase shifts supplement deuteron properties in
characterizing the nuclear interaction. The ratio
$R(x)/R_0(x)$ of the interacting to free correlator has been calculated in
quenched QCD for the following meson and baryon currents~\cite{chu+93}, $J = 
\bar u\gamma_{\mu}d,
\bar u \gamma_{\mu}\gamma_5 d, \bar u\gamma_5 d, \bar u d,
\epsilon_{abc}[c^aC\gamma_{\mu}u^b]\gamma_{\mu}\gamma_5d^c,$
and $\epsilon_{abc}[u^a C\gamma_{\mu}u^b]u^c$.  
Results are consistent with dispersion analysis of e$^+$-e$^-$
and other data in relevant channels, and agree in detail with the channel
dependence expected from the 't Hooft interaction and the instanton liquid model.
Note for subsequent reference, that all correlation functions may be decomposed
into a continuum contribution concentrated near the origin and a resonance
contribution arising from the lowest bound state or resonance which dominates in
the region of 1 fm.  Quenched calculations at $\beta =
6.2$~\cite{hands+95} corroborate the original
$\beta = 5.7$ results. 

\section{Extraction of the instanton content of lattice gluon configurations}
\subsection{Identifying instantons by cooling}

The Feynman path integral for a quantum mechanical problem with degenerate
minima is dominated by paths that fluctuate around stationary solutions to the
classical Euclidean action connecting these minima. In the case of the
double well potential, a typical Feynman path is composed of segments fluctuating
around the left and right minima joined by segments crossing the barrier.  If one
had such a trajectory as an initial condition, one could find the nearest stationary
solution to the classical action numerically by using an iterative local relaxation
algorithm.  In this method, which has come to be known as cooling, one
sequentially minimizes the action locally as a function of the coordinate on each
time slice and iteratively approaches a stationary solution.  In the case of the
double well, the trajectory approaches straight lines in the two minima joined by
kinks and anti-kinks crossing the barrier and the structure of the trajectory can be
characterized by the number and positions of the kinks and anti-kinks.

In QCD, the corresponding classical stationary solutions to the Euclidean action for the
gauge field connecting degenerate minima of the vacuum are instantons, and  the
analogous cooling technique\cite{berg81,teper91} reveals the instantons
corresponding to each gauge field configuration. 

The results of using 25 cooling steps as a filter to extract the instanton content of a
typical gluon configuration are shown in Fig. 1 of
Ref.~\cite{chu+94} using the Wilson action on a $16^3 \times 24$ lattice at
${6}/{g^2} = 5.7$.  As one can see, there is no recognizable structure before
cooling.  Large, short wavelength fluctuations of the order of the lattice spacing
dominate both the action and topological charge density.  After 25 cooling steps,
three instantons and two anti-instantons can be identified clearly.  The action
density peaks are completely correlated in position and shape with the topological
charge density peaks for instantons and with the topological charge density valleys
for anti-instantons.  Note that both the action and topological charge densities are
reduced by more than two orders of magnitude  so that the fluctuations removed by
cooling are several orders of magnitude larger than the topological excitations that
are retained.
%
%

\subsection{The QCD Vacuum}
Several extensive studies of the  instanton content of the SU(3) vacuum
have been performed recently and are reviewed in ref \cite{negele98b}. The results
consistently indicate that the average size of an
instanton when extrapolated to the uncooled vacuum is 0.39 $\pm$.05 fm, in
agreement with the liquid instanton model, and that the topological susceptibility is
$\chi \sim $180 MeV, in agreement with the Veneziano-Witten formula.

\subsection{Comparison of results with all gluons and with only
instantons}\noindent One dramatic indication of the role of instantons in light
hadrons is to compare observables calculated using all gluon contributions with
those obtained using only   the instantons remaining after cooling.  Note that
there are truly dramatic differences in the gluon content before and after cooling. 
Not only has the action density decreased by two orders of magnitude, but also
the string tension has decreased to 27\% of its original value and the Coulombic
and magnetic hyperfine components of the quark-quark potential are essentially
zero.  Hence, for example, the energies and wave functions of charmed and $B$
mesons would be drastically changed.

\begin{figure}[t!] 
\vspace*{-1.0cm}
\begin{center}

\epsfig{file=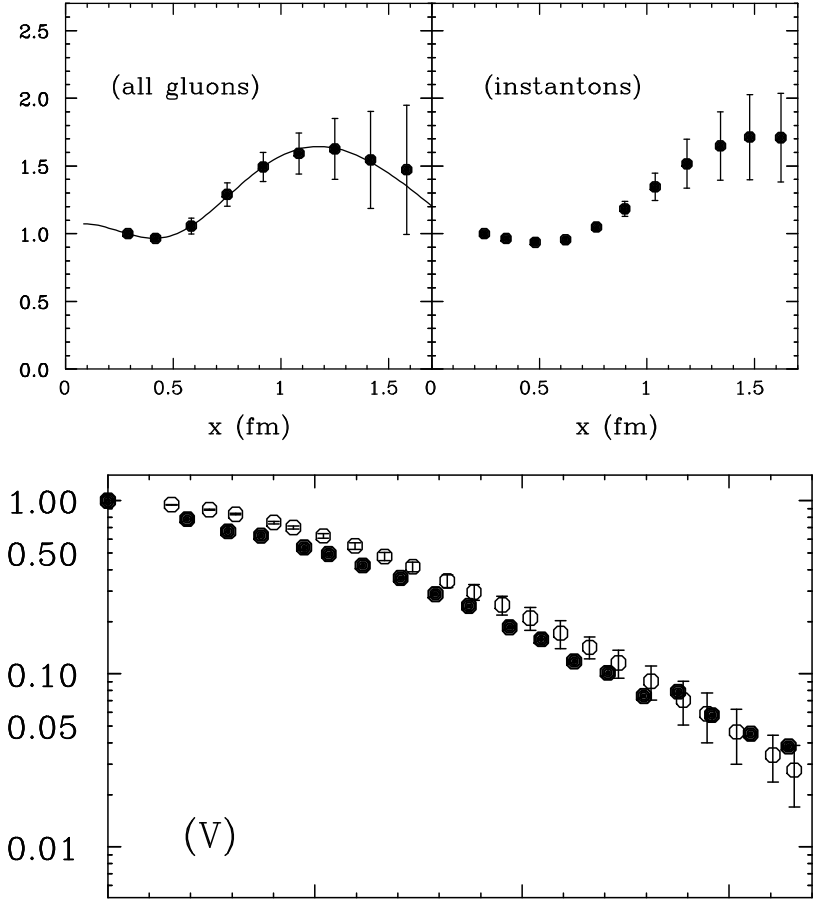}
\caption{Comparison of rho observables calculated with all gluon
configurations and only instantons. The upper left-hand plot shows the vacuum
correlator in the rho channel calculated with all gluons  and the upper right-hand
plot shows the analogous result with only instantons.  The lower plot shows the
ground state density-density correlation function for the rho with all gluons (solid
circles) and with only instantons (open circles). }
\label{F:JN:3}  
\end{center}
\vspace*{-1.0cm}
\end{figure}

As shown in Fig.~\ref{F:JN:3}, however, when the coupling constant, or equivalently,
the lattice spacing, and quark mass are set by the nucleon and pion masses in the
usual way, the properties of the rho meson are virtually unchanged.  The vacuum
correlation function in the rho (vector) channel and the spatial distribution of the
quarks in the rho ground state, given by the ground state density-density correlation
function\cite{chu+91} $\langle \rho| \bar{q} \gamma_0 q(x) \bar{q} \gamma_0 q(0)
| \rho \rangle$, are statistically indistinguishable before and after cooling.  Also, as
shown in Ref.~\cite{chu+94}, the rho mass is unchanged within its 10\% statistical
error.  In addition, the pseudoscalar, nucleon, and delta vacuum correlation functions
and nucleon and pion density-density correlation functions are also qualitatively
unchanged after cooling, except for the removal of the small Coulomb induced cusp
at the origin of the pion. Similarly, the
axial charge matrix elements specifying the spin content of the nucleon,
$\langle \vec{P}\vec{S}|(\bar{q} \gamma^{i} i \gamma_5
q)|\vec{P}\vec{S}\rangle = 2 S^i\Delta q$, are quite similar when calculated
with all gluons and only instantons~\cite{dolgov+98}. 

\section{Quark Zero Modes and their contributions to light
hadrons}\nopagebreak
\subsection{Eigenmodes of the Dirac operator}
In the continuum limit, the Dirac operator for Wilson fermions,\\
$
D \psi_x = \psi_x - \kappa \sum_\mu [ (r-\gamma_\mu) u_{x,\mu}
\psi_{x+\mu} + (r + \gamma_\mu) u^\dagger_{x-\mu,\mu} \psi_{x-\mu} ] $,
approaches the
familiar continuum result 
$
\frac{1}{m} \Bigl[ m+i (\notp + g \, \notA)\Bigr] \psi\  .
$

In the free case, the continuum spectrum is $\frac{1}{m}[m+i |\vec{p}|]$ and the
Wilson lattice operator approximates this spectrum in the physical regime and
pushes the unphysical fermion modes to  very large (real) masses.   In the presence
of an instanton of size $\rho$ at $x=0$, it is shown in Ref.~\cite{ivanenko+98} that the
lattice operator produces a mode with a real eigenvalue which approaches the
continuum result 
$ 
\psi_0(x)_{s,\alpha} = u_{s, \alpha} \frac{\sqrt{2}}{\pi} \frac{\rho}{(x^2 +
\rho^2)^{3/2}}
$
and whose mixing with other modes goes to zero as the lattice volume goes to
infinity.  In addition, instanton-anti-instanton pairs that interact sufficiently form
complex conjugate pairs of eigenvalues that move slightly off the real axis.  Thus,
by observing the Dirac spectrum for a lattice gluon configuration containing a
collection of instantons and anti-instantons, it is possible to identify zero modes
directly in the spectrum.
%

\begin{figure}[t!] 
\vspace{-.5cm}
\begin{center}
\epsfig{file=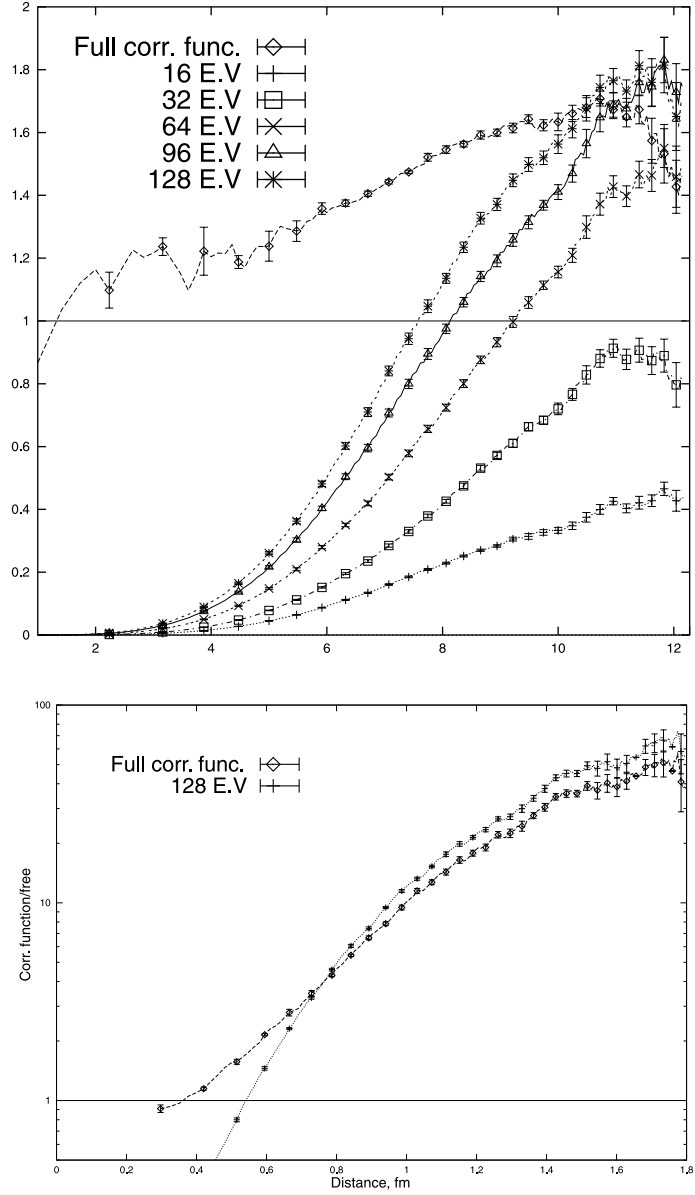}
\caption{Contributions of low Dirac eigenmodes to the vector (upper graph)
and pseudoscalar (lower graph) vacuum correlation functions.  The upper
graph shows the contributions of 16, 32, 64, 96, and 128 eigenmodes compared with
the full correlation function for an unquenched configuration and the lower graph
compares 128 eigenmodes with the full correlation function for a quenched
configuration.}
\label{F:JN:5}  
\end{center}
\vspace{-.5cm}
\end{figure}

Ref.~\cite{ivanenko+98} shows the lowest 64 complex eigenvalues of the Dirac
operator on a $16^4$ unquenched gluon configuration for ${6}/{g^2}=5.5$ and
$\kappa=0.16$, both before and after cooling (where 100 relaxation steps with a
parallel algorithm are comparable to 25 cooling steps).  The cooled plot has
just the structure we expect with a number of isolated instantons with modes on the
real axis and pairs of interacting instantons slightly off the real axis.  However, even
though the uncooled case also contains fluctuations several
orders of magnitude larger than the instantons, it shows
the same structure of isolated instantons and interacting pairs, clearly revealing the
zero modes in the original, uncooled vacuum.

\subsection{Zero mode expansion}\noindent
The Wilson--Dirac operator has the property that $D=\gamma_5 D^\dagger
\gamma_5$, which implies that $\langle \psi_j |\gamma_5| \psi_i \rangle=0$ unless
$\lambda_i = \lambda^*_j$ and we may write the spectral representation of the
propagator
$
\langle x |D^{-1}|y \rangle = \sum_i \frac{\langle x|\psi_i \rangle \langle \psi_{\bar{i}}
|\gamma_5| y\rangle}{\langle \psi_{\bar{i}}|\gamma_5| \psi_i\rangle \lambda_i}
$
where $\lambda_i = \lambda_{\bar{i}}^*$.  A clear indication of the role of zero modes
in light hadron observables is the degree to which truncation of the expansion to the
zero mode zone reproduces the result with the complete propagator.

Fig.~\ref{F:JN:5} shows the result of truncating the vacuum correlation functions
for the vector and pseudoscalar channels to include only low
eigenmodes\cite{ivanenko+98}. On a $16^4$ lattice, the full propagator contains
786,432 modes. The top plot of Fig.~\ref{F:JN:5}  shows the result of including the
lowest 16, 32, 64, 96, and finally 128 modes. Note that the first 64 modes
reproduce most of the strength in the rho resonance peak, and by the time we
include the first 128 modes, all the strength is accounted for.  Similarly, the lower
plot in Fig.~\ref{F:JN:5} shows that the lowest 128 modes also account for  the
analogous pion contribution to the pseudoscalar vacuum correlation function.  
 Similarly, most of the strength of the disconnected
graph contribution to the $\eta^{\prime}$ correlation function, which should be
particularly sensitive to instantons, is already provided by the lowest 32
eigenmodes, and the fermionic definition of the topological charge is nearly
saturated by the lowest 8 modes~\cite{venkataraman+97}.

\subsection{Localization}\noindent
Finally, it is interesting to ask whether the lattice zero mode eigenfunctions are
localized on instantons.  
  This was studied by plotting the quark density distribution for
  individual eigenmodes in
  the $x$-$z$ plane for all values of $y$ and~$t$, and comparing with
  analogous plots of the action density. As expected, for a cooled
  configuration the eigenmodes 
correspond to
linear combinations of localized zero modes at each of the instantons. What is truly
remarkable, however, is
that the eigenfunctions of the uncooled configurations also
exhibit localized peaks at locations at which instantons are identified by
cooling.  Thus, in spite of the fluctuations several orders of magnitude larger
than the instanton fields themselves, the light quarks essentially average
out these fluctuations and produce localized peaks at the topological
excitations.  When one analyzes a number of eigenfunctions, one finds that
all the instantons remaining after cooling correspond to localized quark
fermion peaks in some eigenfunctions.  However, some fermion peaks are
present for the initial gluon configurations that do not correspond to
instantons that survive cooling.  These presumably correspond to
instanton--anti-instanton pairs that were annihilated during cooling.

\section{Conclusion}\noindent
Altogether, the lattice calculations reported here provide strong evidence
that instantons play a dominant role in quark propagation in the vacuum
and in light hadron structure.  We have shown that the instanton content of
gluon configurations can be extracted by cooling, and that the instanton size is
consistent with the instanton liquid model and the topological susceptibility agrees
with the Veneziano-Witten formula. We obtain striking agreement between vacuum
correlation functions, ground state density-density correlation functions, and masses
calculated with all gluons and with only instantons.  Zero modes associated with
instantons are clearly evident in the Dirac spectrum, and  account for the $\rho, \pi$,
and $\eta\prime$ contributions to vacuum
correlation functions.  Finally, we have observed directly quark localization at
instantons in uncooled configurations.

\subsubsection*{Acknowledgments}\noindent
It is a pleasure to acknowledge the essential role of Richard Brower, Ming Chu,
Dmitri Dolgov, Jeff Grandy, Suzhou Huang, Taras Ivananko, Kostas Orginos, and
Andrew Pochinsky who collaborated in various aspects of this work.  We are also
grateful for the donation by Sun Microsystems of the 24 Gflops E5000 SMP cluster
on which the most recent calculations were performed and the computer resources
provided by NERSC with which this work was begun.



\end{document}